\documentclass[usenatbib]{basi}
\usepackage[T1]{fontenc}
\usepackage[british]{babel}
\usepackage[varg]{txfonts}
\usepackage{rotating}
\usepackage{dcolumn}

\begin{document}
\title[Radio studies of novae]{Radio studies of novae: a current status report and highlights of new results}
\author[Roy et al.]
{\parbox{\textwidth}{
Nirupam~Roy$^1$\thanks{email: \texttt{nroy@nrao.edu}}, Laura~Chomiuk$^{1, 2}$, Jennifer~L.~Sokoloski$^3$, Jennifer~Weston$^3$, Michael~P.~Rupen$^1$,  Traci Johnson$^{1, 4}$, Miriam~I.~Krauss$^1$, Thomas~Nelson$^5$, Koji~Mukai$^{6, 7}$, Amy~Mioduszewski$^1$, Michael~F.~Bode$^8$, Stewart~P.~S.~Eyres$^9$, and Tim~J.~O'Brien$^{10}$}
\vspace{0.4cm}
\\
\parbox{\textwidth}{
$^1$ National Radio Astronomy Observatory, P.O. Box O, Socorro, NM 87801 USA\\
$^2$ Department of Physics and Astronomy, Michigan State University, East Lansing, MI 48824\\
$^3$ Columbia Astrophysics Laboratory, Columbia University, New York, NY 10027, USA\\
$^4$ Physics and Astronomy Department, Carleton College, 1 N. College St., Northfield, MN 55057\\
$^5$ School of Physics and Astronomy, University of Minnesota, 116 Church St. SE, Minneapolis, MN 55455\\
$^6$ CRESST and X-ray Astrophysics Laboratory, NASA/GSFC, Greenbelt, MD 20771, USA\\
$^7$ Center for Space Science and Tech., Univ. of Maryland Baltimore County, Baltimore, MD 21250 USA\\
$^8$ Astrophysics Research Institute, Liverpool John Moores University, Birkenhead, CH41 1LD, UK\\
$^9$ Jeremiah Horrocks Institute, University of Central Lancashire, Preston, PR1 2HE, UK\\
$^{10}$ Jodrell Bank Centre for Astrophysics, University of Manchester, Manchester M13 9PL, UK
}}

\pubyear{2012}
\volume{40}
\pagerange{\pageref{firstpage}--\pageref{lastpage}}

\date{Received 2012 August 16; accepted August 30}

\maketitle
\label{firstpage}

\begin{abstract}
Novae, which are the sudden visual brightening triggered by runaway 
thermonuclear burning on the surface of an accreting white dwarf, are fairly 
common and bright events. Despite their astronomical significance as nearby 
laboratories for the study of nuclear burning and accretion phenomena, many 
aspects of these common stellar explosions are observationally not 
well-constrained and remain poorly understood. Radio observations, modeling 
and interpretation can potentially play a crucial role in addressing some of 
these puzzling issues. In this review on radio studies of novae, we focus on 
the possibility of testing and improving the nova models with radio 
observations, and present a current status report on the progress in both the 
observational front and theoretical developments. We specifically address the 
issues of accurate estimation of ejecta mass, multi-phase and complex ejection 
phenomena, and the effect of a dense environment around novae. With highlights 
of new observational results, we illustrate how radio observations can shed 
light on some of these long-standing puzzles. 
\end{abstract}


\section{Introduction}\label{nova:intro}

Novae occur in binary stellar systems where mass is transferred onto a white 
dwarf (WD) from either a main sequence, subgiant, or red giant companion. As 
the accreted material builds up on the WD surface, the temperature of the 
degenerate Fermi gas increases without any change of pressure. Ultimately, a 
thermonuclear runaway (TNR) ensues and subsequently the critical temperature 
is reached, breaking the degeneracy of the accreted layer and resulting in 
ejection of mass from the WD surface (see e.g. \citealt{sta08}, also 
Starrfield et al.~2012, this Volume). This event is observed as a nova --- a sudden 
visual brightening at optical wavelengths. For more on the optical studies of 
novae, see the reviews by Anupama \& Kamath (2012) and Shore (2012) in this 
Volume. Nova events are fairly common \citep[$\sim$35 novae per year in the 
Milky Way;][]{sha97} and bright (sometimes even visible to the naked eye for 
days to weeks). Thus, novae provide valuable nearby laboratories for the study 
of nuclear burning and accretion phenomena. Additionally, if the ejected mass 
is less than that accreted since the last outburst, then nova-hosting WDs will 
grow in mass with time. This makes them interesting candidates for progenitors 
of Type Ia supernovae (SNe Ia). See \citet{sta04}, \citet{del96} and also 
Kato \& Hachisu (2012; this Volume) for more details on the connection between 
novae and SN Ia progenitors. 

Novae have been observed in detail for the greater part of the 20th century 
\citep[e.g.,][]{pg64}, and the basics of nova theory have been established 
since the 1970's  \citep[e.g.,][]{sta72}. One might therefore expect that we 
understand these common stellar explosions in depth, but many fundamentals 
remain poorly understood or observationally unconstrained. Some intriguing 
discrepancies exist between observations and theoretical predictions --- like 
ejecta masses which are observed to be an order of magnitude larger than 
predicted by theory. In addition, observations indicate that nova explosions 
are complex, with multiple phases --- and perhaps physical drivers --- of mass 
ejection \citep{wil12}. Models agree that material is not ejected from the WD 
surface in a single impulsive burst \citep{pri86}, but the consistency of 
observed and predicted ejection histories has not been extensively tested. 
Radio observations are ideal for pursuing these issues, because they trace the 
majority of the ejected mass in a relatively simple and easily modeled fashion. 

In this review, we summarize our current state of understanding of the radio 
emission mechanism and evolution of novae, and focus on how radio observations 
can test nova models. In Section 2 we describe the standard model for radio 
light curves (thermal bremsstrahlung from an expanding spherical shell). In 
Section 3 we discuss estimates of ejecta mass from radio observations, and how 
models may be refined to produce more accurate ejecta masses. In Section 4, we 
show some recent evidence that the ejection of material from novae is 
multi-phase and complex, and illustrate how radio observations can shed light 
on long-standing observational puzzles at other wavelengths while testing nova 
theory. In Section 5, we discuss some of the observational effects of a dense 
circumbinary environment around novae. Finally, Section 6 contains a few 
concluding remarks. 

\section{Radio emission from novae}\label{nova:radio}

The radio emission from novae is typically much longer lasting than the 
optical emission, evolving on timescales of years rather than months. Radio 
observations at a range of epochs yield information on different 
characteristics of the nova outburst, from the distance to the system at very 
early times, to the mass of the ejecta as the light curve evolves. Therefore, 
an observing strategy that monitors novae at both early and late times is 
crucial to gain a full understanding of their properties.

\subsection{``Standard model'' of expanding thermal ejecta}\label{nova:em10}

Radio emission from novae was first detected by \citet{hje70} for HR~Del and 
FH~Ser. Both sources were observed to have steep positive spectral indices 
($\alpha > 0$, $f_\nu \propto \nu^\alpha$) during the early-time radio light 
curve, when radio luminosity is increasing with time. This signal was 
interpreted as optically-thick thermal emission, and showed a brightness 
temperature comparable to the typical kinetic temperature of electrons in 
photo-ionized plasma ($T_b \approx 10^4$ K). At late times, when the light 
curves decline, the spectral index is almost flat ($\alpha \approx -0.1$). 
Later, these properties were found to be general characteristics of most novae 
with detectable radio emission.

Thermal bremsstrahlung from the warm ejecta is believed to be the primary 
mechanism of radio emission from classical novae. There are exceptions like 
GK Per or RS Oph with significant synchrotron emission in the radio 
\citep{sea89,anu05,rup08,sok08}; these are thought to be explosions expanding 
into unusually dense environs. Overall, non-thermal radio emission from 
classical novae appears to be rare \citep{bod87}.

At the zeroth order, the thermal bremsstrahlung emission is simple to model, 
and, with some reasonable assumptions and/or complementary observations, it 
can be used to derive physical parameters like ejected mass. Such a simple 
model of thermal emission from an expanding shell of plasma has been used to 
explain radio light curves from $\sim$10 novae in the past 
\citep[e.g.][]{sea77,hje79,sea80,kwo83}.

The recent review of radio emission from novae by \citet{sea08} 
comprehensively treats the theory of thermal bremsstrahlung emission from nova 
shells, and predicted radio light curves. In brief, the model assumes a 
spherically symmetric isothermal shell of ionized gas with a power law density 
gradient (radial profile of the number density $n(r) \sim r^p$ where $p=2-3$ 
is found to be a good fit to most of the observations). The time evolution of 
the system is introduced through a kinematic model of the expansion of the 
shell. Different physical models lead to differences in the density profile at 
relatively small radii, therefore affecting the predicted radio light curves 
at late times.

\begin{figure}[tbp]
\centering
\includegraphics[width=12cm, angle=0]{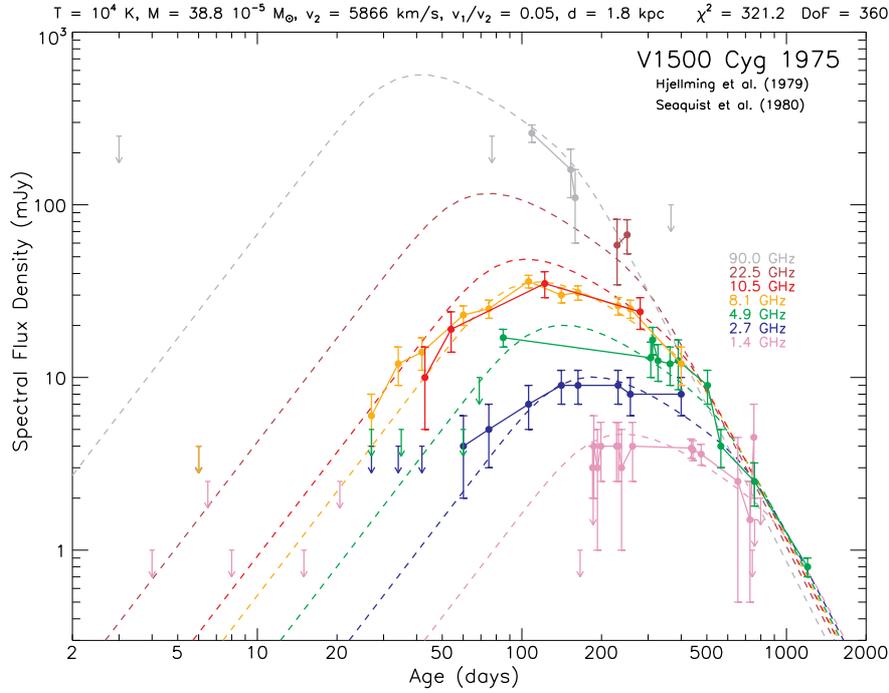}
\caption{{\it The multi-frequency radio light curve of V1500~Cyg (from 
\citealt{hje79} and \citealt{sea80}), fit by the simple Hubble flow model 
(dashed lines). The model light curves for 1.4 - 90 GHz matches with the 
observed values within measurement uncertainties and give reasonable values 
for the physical parameters like ejecta mass and velocity.}}
\label{hubble}
\end{figure}

For the simplest of these models, known as the ``Hubble flow'' model, the 
ejection is instantaneous, the outflow speed increases linearly with radius, 
and the same amount of mass is expelled at all velocities, implying $p=2$ 
\citep{sea77,hje79,sea80,hje96}. The shell is expelled with a range of 
velocities (e.g., $v_{\rm min}/v_{\rm max} = 0.05$ in Figure \ref{hubble}), so 
the shell has a hard inner edge.

In contrast, the ``variable wind'' model assumes continuous mass loss over an 
extended time period, significantly longer than the timescale of the radio 
light curve \citep{kwo83,hje90}. At all times, mass continues to refill the 
density profile at small radius, so there is no evacuated cavity at the center 
of the ejected shell (unlike in the Hubble Flow model). It is important to 
note that, at least for some novae, there is observational evidence of 
prolonged periods of ejection \citep[e.g.][]{gal78}. Again, for simplicity 
$p=2$ is often assumed in the variable wind model, implying a constant 
$\dot{M}_{w}/v_{w}$ over time ($p=2$ is not required by theory of nova winds, 
but it can be easily solved analytically, while $p=3$ requires numerical 
integration; see \citealt{sea80} for a comparison of light curves with $p=2$ 
and $p=3$). 

An intuitive tweak to the variable wind model is a wind that varies, and then 
ceases. In this case, the nova ejecta will detach from the WD and leave a 
cavity at small radius. This ``unified'' model essentially combines the 
variable wind model (at early times) and the Hubble flow model (at late times) 
to explain the radio light curve of V1974 Cyg, a nova with one of the highest 
quality radio light curves ever obtained \citep{hje96}.

For a given kinetic temperature of the plasma, outflow velocity, velocity 
gradient, total ejected mass, and distance, the Hubble flow model predicts the 
temporal and spectral behaviour of the radio light curve. It predicts a $t^2$ 
rise and a steep spectral index ($\alpha = 2$) for the initial optically thick 
phase, and a $t^{-3}$ decay with a flat spectral index ($\alpha = -0.1$) for 
the late time optically thin phase. At intermediate times, when the emission 
is transitioning from optically thick to optically thin and the radio 
photosphere is receding through the ejecta, the model predicts  $t^{-4/3}$ 
decline with $\alpha = 0.6$. Historically, this model has been successful in 
explaining the radio evolution for a number of sources over a range of 
frequencies and timescales. The observed $\sim t^{-3}$ decay at late times 
indicates the importance of an inner boundary to the thermal shell; the 
variable wind model can not reproduce the radio light curve at all time 
scales. However, a prolonged wind model best explains the very early time 
optical emission from novae, interpreted as being produced before the ejecta 
detach from the star --- hence, the need for the unified model of 
\citet{hje96}. Also, please note that all of these scenario are simplified 
models, and, in the words of \citet{sea08}, they are not {\it ``firmly rooted 
in a detailed understanding of the mass-loss mechanism, but they do constitute 
at least an initial framework for interpreting the radio data, and for 
obtaining insight into this mechanism''}.

Figure \ref{hubble} shows an example of the best fit Hubble flow model for the 
multi-frequency radio light curve of V1500~Cyg. The data used here are from 
\citet{hje79} and \citet{sea80}. The model provides reasonable values for 
physical parameters like ejecta mass (M$_{\rm ej} = 4 \times 10^{-4}$ 
M$_{\odot}$) and velocity (v$_{\rm max} = 6000$ km s$^{-1}$) for this nova, 
and provides a good representation of the observed light curves over 1.4 - 90 
GHz. However, please note that the measurement uncertainties for the flux 
densities are quite large here, and hence it would be difficult to notice 
deviations from the standard model.

\subsection{Role of the Karl G. Jansky Very Large Array and of the ``E-Nova 
Project''}\label{nova:em101}

Though these models of expanding thermal ejecta were able to fit the 
multi-frequency radio light curves of a number of novae (e.g., HR~Del, FH~Ser, 
V1500~Cyg; \citealt{hje79}), there were clear indications of deviations from 
the simple model in many cases (e.g. \citealt{tay87,hje96,llo96}). Recent work 
(Johnson et al., in preparation), based on careful review and some reanalysis of 
the ``historical'' radio data, shows that such deviations are in fact generic 
features in the radio light curves of novae. Such deviations may require one 
or more modifications to the simple model, such as (i) geometrical complexity, 
e.g., non-spherical ejecta and clumpy small scale structure; (ii) spatial 
and/or temporal variation of plasma temperature; (iii) multiple episodes of 
mass ejection; and (iv) interaction with surrounding material giving rise to 
additional thermal and/or non-thermal components.

The Karl G. Jansky Very Large Array (VLA) is now playing a crucial role in 
this context. The upgraded new receivers, backend, and correlator provide 
dramatic improvements in frequency coverage, instantaneous bandwidth, and 
sensitivity (for a summary of the upgrade, see \citealt{per11}). The improved 
sensitivity implies that we are now able to detect the weak radio emission 
expected within a few days of the optical outburst, so we are observing novae 
earlier than ever before. We can carry out measurements over a wide range of 
frequencies ($1 - 40$ GHz) typically in $1.5 - 2.5$ hours of observing time, 
and constrain the spectral index of radio emission accurately. This 
spectacular new facility is enabling much more detailed radio measurements of 
Galactic novae than ever before, and the effort to obtain these exquisite new 
data is spear-headed by our E-Nova team. 

Over the last 2 years, our team has obtained VLA observations of 10 novae 
within one month of discovery, sometimes as early as 3 days after the onset of 
the outburst. Many of these observations have resulted in non-detections 
\citep{v5587,oph2012,sgr2012}, but several novae have been detected early 
\citep[e.g.,][]{v407,v1723,v55880,v55881,v55882,asco2012}. Our monitoring 
program has led us to carry out long term (more than one year) monitoring 
campaigns on three novae: V407~Cyg, V1723~Aql, and T~Pyx, and we are presently 
beginning monitoring of several younger targets. Essentially, we plan to 
target new Galactic novae visible to the VLA which are optically bright ($V < 
8$ mag, a selection criterion matched by the \emph{Swift} nova group, 
promising complementary X-ray and UV/optical observations) and/or which show 
unusual interesting behaviour at other wavelengths, capturing the attention of 
the broader nova community. After several epochs of VLA observations, we 
evaluate the strength of radio detection and decide if further follow-up 
observations are warranted.

In addition to acquiring radio light curves, multi-wavelength partnerships 
are a critical component of the ``E-Nova Project''. Our goal is to obtain 
spatially-resolved radio images (with eMERLIN, VLBA, EVN, and VLA extended 
configuration), millimetre observations (with SMA, CARMA, and ALMA), X-ray 
photometry and spectroscopy (with \emph{XMM, Suzaku, Swift}), and 
high-resolution optical spectroscopy (with FLWO 1.5m, SOAR, and Magellan) for 
our targets. We also invest significant effort in modeling these observations, 
to thereby extend and constrain the current models of novae by confronting 
them with the widest possible range of consistent, high-quality, 
multi-wavelength data. For some of the details of our VLA observations and 
results, see \citet{kra11} and \citet{cho12}. Also, see Kantharia (2012, this Volume) for a status report and interesting results from lower 
frequency radio observations of novae using the GMRT.

\section{Nova ejecta masses from radio light curves}

One of the most exciting prospects for radio observations of novae is the 
accurate measurement of ejected masses. As we have seen in the previous 
section, the radio emission from most novae arises predominantly from the 
thermal ejecta, and as the radio photosphere gradually recedes through the 
ejecta it samples the entire mass profile. The mass of ejecta is a fundamental 
prediction of nova models, and thereby provides a direct test of nova theory.

\begin{figure}[tbp]
\vspace{-1.5cm}
\centering
\includegraphics[width=11cm, angle=90]{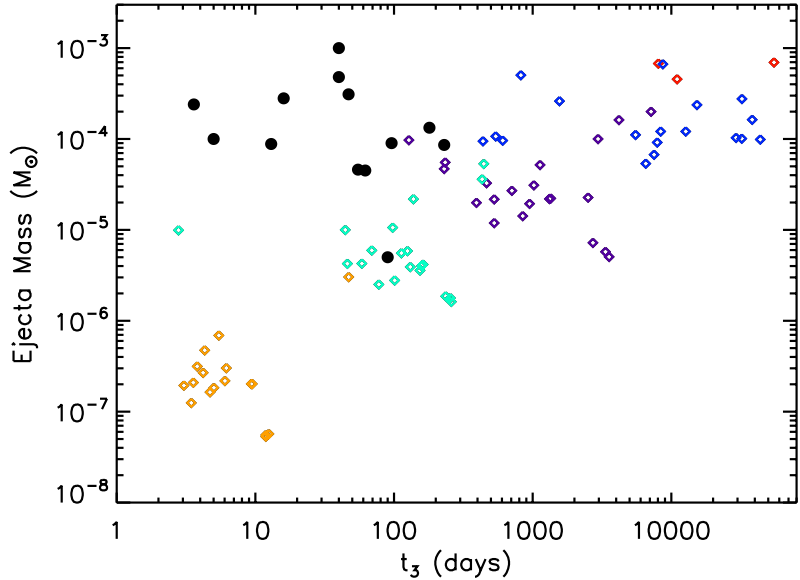}
\vspace{-1.9cm}
\caption{{\it Nova ejecta mass plotted against the time for the optical light 
curve to decline by three magnitudes ($t_3$). Theoretical predictions from 
\citet{yar05} are shown as open diamonds, and color-coded by the mass of the 
WD that hosts the explosion (red = 0.4 M$_{\odot}$, blue = 0.65 M$_{\odot}$, 
purple = 1.0 M$_{\odot}$, cyan = 1.25 M$_{\odot}$, orange = 1.4 M$_{\odot}$). 
Observational estimates from radio data are plotted as filled black circles, 
as compiled by \citet{sea08} and Johnson et al.~(in preparation).}}
\label{mej_t3}
\end{figure}

\subsection{Testing nova models with observed ejecta masses}\label{nova:test}

In general, the WD properties, like the WD mass, internal temperature, and 
accretion rate from the companion star, should dictate the fundamentals of the 
novae explosions, like recurrence time, ejecta mass, and explosion energetics. 
Theory predicts simple relationships between the WD properties and explosion 
characteristics \citep{yar05}, but these predicted relationships have been 
difficult to test observationally. The internal WD temperature is practically 
impossible to measure, as it is buried deep under the complexities of the 
accreting WD's surface. In classical novae, the host binary systems are 
usually not known before nova outburst, and therefore the only hope of 
constraining WD mass and accretion rate is to study the binary after it 
returns to quiescence. However, many nova-host binaries are intrinsically very 
faint in quiescence, and the accretion rate immediately post-nova may not be 
representative of the average accretion rate \citep[e.g.,][]{sha86}. In the 
realm of fundamental explosion properties, we have seen that radio data can 
provide measurements of the ejecta mass. The luminosity and energetics of the 
nova explosion should also be measurable, with a suite of multi-wavelength 
data. Therefore, our best hope for testing theory in a significant sample of 
novae is to cross-compare different properties of nova explosions, to see if 
they correlate as predicted and populate the expected parameter space.

It has been pointed out for some time that there is a discrepancy between 
observed and predicted ejecta masses in novae, where the observed masses are 
roughly an order of magnitude greater than predicted \citep{sta98,geh02}. This 
discrepancy is illustrated in Figure \ref{mej_t3}, which shows that 
practically all observed novae have ejecta masses in the fairly narrow range 
of few $\times 10^{-5}$ to few $\times 10^{-4}$ M$_{\odot}$, approximately the 
maximum ejecta mass predicted for \emph{any} parameters in the \citet{yar05} 
models. In Figure \ref{mej_t3}, we plot ejecta mass against optical decline 
time, and expect these two parameters to be correlated (as in the Yaron et 
al.~models), because the optical photosphere should recede more quickly 
through less massive ejecta. We do not see any such correlation for the 
observed novae. The discrepancy between theory and observation is largely due 
to the fact that the observed novae have relatively fast optical decline times 
($t_3 \lesssim 100$ days), which should translate into relatively small ejecta 
masses ($10^{-7}-10^{-5}$ M$_{\odot}$). Few slow or very slow novae have been 
studied in the radio, to test if the discrepancy persists for long $t_3$.

Potential physical causes of the discrepancy are poorly understood, although 
mixing processes are a candidate culprit. \citet{sta98} suggest that  masses 
of accreted material on the WD could theoretically grow larger if heavy 
elements (i.e., C/O or O/Ne) are not mixed into the H layer. A lack of mixing 
would minimize the opacity in the accreted layer, maximize cooling, and 
maximize the duration of accretion before a thermonuclear runaway ensues. Such 
a lack of heavy-element mixing might be accomplished if there exists a He-rich 
buffer layer between the newly-accreted H-rich surface layer on the WD and the 
heavier-element-dominated WD itself. Of course, this possible explanation is 
quite speculative, and there are in fact contrasting results from recent 
numerical simulations \citep[e.g.,][]{cas11,gla11}, Clearly, significantly 
more work --- on both observational and theoretical sides --- is required to 
resolve the ejecta mass discrepancy.

\subsection{Clumpy nova ejecta can masquerade as massive nova ejecta}

On the observational side, several complexities currently limit the accuracy 
of ejecta mass determinations, with the most significant being clumping in the 
ejecta. Since the thermal radio emission is directly proportional to the 
square of the electron density, any clumping will increase the radio flux 
density and the derived ejecta mass. Volume filling factors of $f = 10^{-1} - 
10^{-5}$ are estimated in nova ejecta \citep[e.g.,][]{sai94,mas05,ede06,sha12}, 
which can boost the radio luminosity by corresponding factors of 5 -- 2000 
above what would be emitted for unity filling factor (\citealt{abb81,hey04}; 
assuming that all of the emitting material is in the high-density clumps). 
Current radio models of clumping in novae are relatively simple, assuming a 
constant filling factor and temperature throughout the ejecta. This simple 
models scale the radio light curve to be brighter, but do not change the time 
evolution of the radio light curve.

Clumping was invoked as an explanation of the observed high ejecta mass for 
V723~Cas by \citet{hey05}. The best fit Hubble flow model to explain the radio 
flux densities from the MERLIN observations for this source gave an ejecta 
mass of $1.13 \times 10^{-4} M_\odot$, significantly higher than theoretical 
predictions. Based on this discrepancy, they argued that the ejecta is likely 
to be clumpy, and hence the actual ejecta mass is lower. This argument was 
also supported by the irregularities visible on the resolved image of the 
shell (though much of these was later attributed to instrumental effect due to 
limited $uv$ coverage of the MERLIN observations; \citealt{hey07}). We are 
considering clumping in our interpretation of the recently-obtained VLA light 
curve for the recurrent nova T~Pyx (Nelson et al.~2012d). The first 
nova outburst from T~Pyx was detected in 1890, and additional events were 
observed in 1902, 1920, 1944, 1967, and most recently 2011. Clumping is 
expected to be important, because T~Pyx is surrounded by an H$\alpha$+[N~II] 
nebula which is interpreted as the ejecta from previous nova outbursts, and 
which clearly displays a clumpy morphology \citep{shar97}. Our analysis 
indicates that including clumping in the model may be crucial to explain the 
radio light curve of T~Pyx (Nelson et al.~2012d).

Although some of the discrepancy between predicted and observed ejecta masses 
could be due to clumping and temperature variations, in actuality a wide range 
of techniques using diverse wavebands find similar ejecta masses, and they are 
consistently higher than predicted \citep[e.g.,][]{sta98,sch02,geh02}. In 
addition, hints from optical spectroscopy imply the existence of significant 
reservoirs of gas which will not emit at radio wavelengths, either because 
they are neutral or very hot \citep{wil94,fer98}. Therefore, most published 
estimates of ejecta mass are likely lower limits and the discrepancy may be 
worse than it appears in Figure \ref{mej_t3}.

\subsection{Ejecta masses in recurrent novae}

Recurrent novae (binary systems that have been recorded to host nova outbursts 
more than once in human history) provide a wealth of information, as compared 
with classical novae. In recurrent novae, we know the recurrence time between 
outburst, and are aware of the binary system over a long time baseline so that 
significant effort can be invested in measuring the WD mass and accretion rate 
during quiescence (e.g., T Pyx; \citealt{pat98,sel08,uth10}). Therefore, 
recurrent novae can provide some of the most stringent and thorough tests of 
nova theory, allowing us to compare explosion mass and energetics against 
other fundamental properties of the binary system. 

One of the most important predictions of nova models has been that most novae 
lead to a net loss in mass from accreting WDs, implying severe difficulties in 
growing WDs to the Chandrasekhar mass so that they explode as SNe Ia. Only 
less energetic novae outburst, which take place on massive WDs with high 
accretion rates, should eject less mass than they had accreted since the last 
nova, allowing them to slowly grow (\citealt{yar05}; but also see 
\citealt{sta12a} who recently find that WDs grow in mass for a range of 
outburst parameters). This is also the same class of novae that have 
relatively short quiescent intervals between outbursts, and will likely be 
recognized as recurrent novae. Recurrent novae are one of the only tests of 
whether --- and in what circumstances --- accreting WDs can be SN Ia 
progenitors, by measuring which is larger: the mass ejected in the nova or the 
accretion rate times the recurrence time. However, the models of \citet{yar05} 
find that the difference between the accreted and ejected mass is seldom 
greater than a factor of $\sim$2. Therefore to truly constrain if a WD is 
growing or shrinking in mass, we need accurate and precise measurements of 
both the accretion rate and the ejecta mass. Radio observations are key in 
determining the latter, but unfortunately our interpretation of these data 
does not yet yield ejecta masses which are sufficiently accurate. Clearly, 
future work is needed. In this context, with our ongoing analysis of the radio 
light curve of T Pyx, in near future we will be able to compare the mass of 
the ejecta in that system to the mass accreted in quiescence (Nelson et 
al.~2012d).

\subsection{Future Prospects for understanding the mass discrepancy}

Radio recombination lines (RRLs) of hydrogen have not yet been detected in 
novae, but they have the potential to shed light on clumping in ejecta, and 
thereby nova ejecta masses. With measurements of RRLs at a range of 
frequencies, we could extract the filling factor as a function of radius, 
along with constraints on density and temperature 
\citep[e.g.,][]{roe91,ana00}. For a significant RRL detection at a few $\times 
10$ GHz with the upgraded VLA, we estimate that a nova should have a flux 
density $\gtrsim$10 mJy and be at least partially optically thin; such flux 
densities are regularly reached by Galactic novae, so the future looks bright 
for studies of RRLs in novae.

No matter how accurate our modeling of radio light curves becomes, we must 
also pursue a multi-wavelength strategy that prioritizes optical spectroscopy, 
and ideally UV and infrared spectroscopy. Combined studies which take 
advantage of synoptic spectroscopy, multi-frequency radio light curves, and 
detailed modeling of multi-wavelength data with CLOUDY hold the most promise 
for accurate estimates of ejecta mass.

In addition, to fairly characterize the offset between observed and 
theoretical ejecta masses, and ensure that the suggested discrepancy is not 
dominated by outliers, we need to constrain the \emph{distribution} of 
observed nova ejecta masses and compare them with the predicted 
\emph{distribution}. We require mass estimates --- or at least limits on the 
mass --- for a sample of novae representative of the Milky Way population. 
Currently, targets for multi-wavelength follow-up are rarely chosen because 
they are representative, but instead because they are particularly interesting 
in some waveband or because they are optically bright. This has likely biased 
our samples of novae in difficult-to-quantify fashion. In addition, radio 
non-detections of novae have rarely been published, leading to a bias where 
only the most massive novae are permitted onto Figure \ref{mej_t3}. At 
present, it is difficult to assess if the paucity of novae with observed 
ejecta masses $<10^{-5}$ M$_{\odot}$ is due to this bias, or if low-mass 
ejecting novae really do exist in smaller numbers than predicted by theory. To 
make progress on this question and test if observed novae are consistent with 
theoretical predictions, we require an observing philosophy that prioritizes 
completeness, understanding of selection effects, and inclusion of censored 
data. Future radio transient surveys of the Galactic Bulge, like ThunderKAT 
with MeerKAT, should make significant progress on our understanding of the 
discrepancy between observed and predicted ejecta mass. 

\section{Complex multi-phase ejections traced by radio light curves}

All models of the radio emission from novae presented in this review assume a 
smooth homologous flow with velocity gradient for the ejecta that has the 
fastest material on the outside, and slowest material in the interior of the 
shell. However, complementary observations at other wavelengths illustrate the 
complexity of mass ejection in novae. In particular, hard X-ray emission is 
detected in a growing number of novae at early times, and has been interpreted 
as evidence of fast material ejected at late times sweeping up, and shock 
heating earlier, slower ejecta to X-ray emitting temperatures - the exact {\it 
opposite} of the velocity gradient in the radio models 
\citep{obr94,muk01,muk08}. From optical spectroscopy also, it was suggested 
that the velocity of the nova wind may increase with time, so that material 
expelled at later times catches up with earlier-expelled material and produces 
shocks \citep{fri87, war08}. A two phase wind mass loss model was proposed by 
\citet{kwo83} to explain the optical and radio data for a few novae. In 
addition, many features of nova optical light curves remain poorly understood. 
While some light curves simply decline as expected for a receding photosphere 
in expanding ejecta, others show plateaus that can last from days to months 
\citep{kat11}, or multiple secondary maxima \citep[e.g.,][]{mun08,hou10} --- 
consistent with ``stalled-out'' or growing optical photospheres. Radio 
observations can shed light on the mechanisms and history of mass ejection in 
novae, and on these puzzling multi-wavelength phenomena, because they are an 
excellent probe of the density profile of the ejecta.

\begin{figure}[tbp]
\centering
\includegraphics[width=11cm, height=12cm, angle=0]{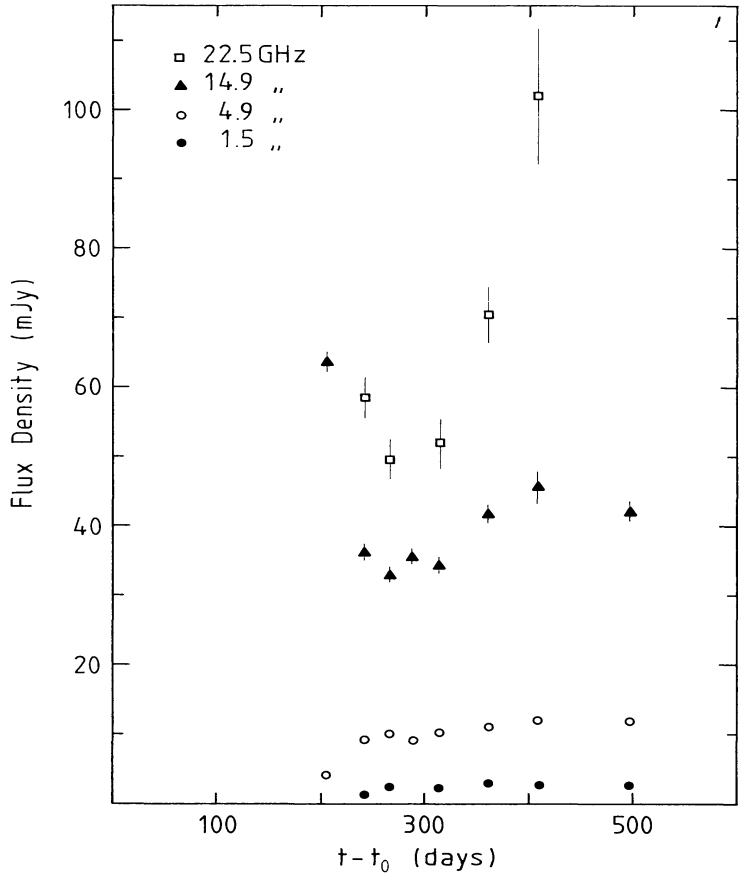}
\vspace{-0.5cm}
\caption{{\it VLA radio light curves of QU~Vul spanning 1.5 - 22.5 GHz 
\citep{tay87}. In the first few epochs, the higher frequencies are anomalously 
bright, but they subsequently fade to participate in the evolution expected 
from the standard model of expanding thermal ejecta.}}
\label{quvul}
\end{figure}

\begin{figure}[tbp]
\centering
\includegraphics[width=12.5cm, height=9.5cm, angle=0]{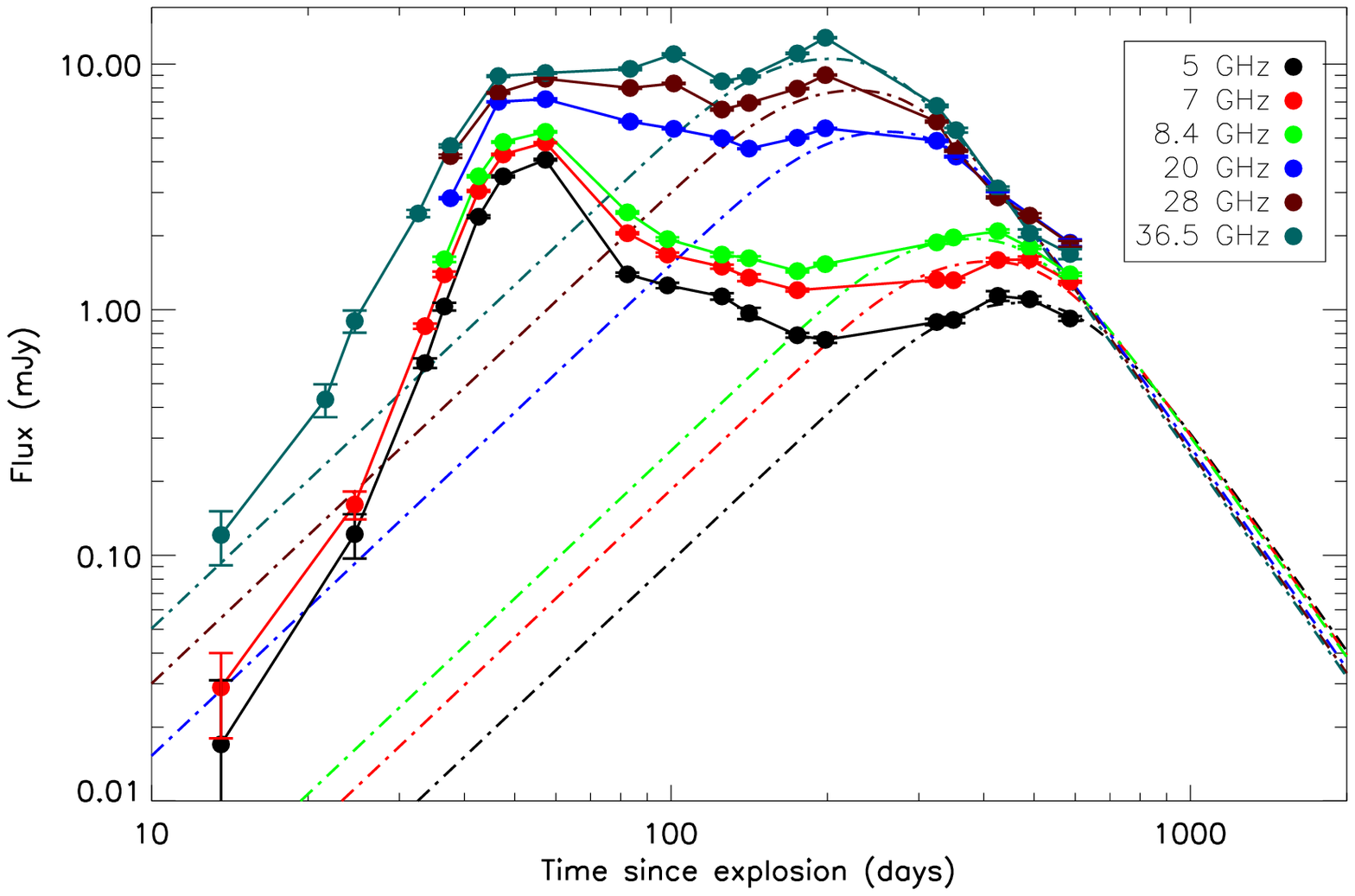}
\vspace{-0.5cm}
\caption{{\it VLA radio light curves of V1723~Aql spanning 5 - 40 GHz. The 
light curves have a rapid initial rise ($f_{\nu} \propto t^{3.3}$) to an 
anomalous peak around Day 50 with a nearly flat spectrum. The best fit Hubble 
flow model shows that late-time data probably adhere to the standard model of 
expanding thermal ejecta.}}
\label{v1723aql}
\end{figure}

\subsection{Early time bumps in radio light curves}

The first clear example of an early-time anomalous maximum in a nova radio 
light was the 1984 explosion of QU Vul (\citealt{tay87}; Figure \ref{quvul}). 
Regrettably, the first epoch of radio observation was not until 206 days after 
discovery, but it revealed a very high flux density at 15 GHz (63 mJy) and a 
very steep spectral index of $\alpha = 2.4$. Over the next $\sim$50 days, the 
15/22 GHz light curves declined while 1.5/4.9 GHz light curves gently rose, 
leading to a flattening of the spectral index. Late time evolution of QU Vul 
was roughly consistent with the standard model of expanding thermal ejecta, 
showing a secondary maximum at high frequencies and continue rise at low 
frequencies. \citet{tay87} show that the early time bump could be the 
signature of a multi-phase ejection, wherein a shell is expelled at early 
times, followed by a faster wind which shocks this shell. In their model, the 
velocity differential between the shell and wind is of order $\sim$200 km 
s$^{-1}$, and the shocked material emits optically thin thermal emission at 
radio wavelengths. However, this shock is embedded in the ionized ejecta, 
which provides an absorbing screen at radio wavelengths and leads to the 
observed $\alpha = 2.4$.

More recently, our extensive monitoring of the classical nova V1723~Aql also 
shows unusual behaviour and significant departure from the Hubble flow model 
predictions \citep{kra11}. The VLA multi-frequency light curve for this source 
is shown in Figure \ref{v1723aql}. There is a bump with very fast ($\sim 
t^{3.3}$) rise. Unlike in QU Vul, as the flux density rises to this early time 
bump, the spectral index \emph{flattens} from $\alpha = 1.1$ to $\alpha = 0.4$ 
around the Day $\sim$50 maximum. Subsequently, the flux densities at lower 
frequencies decrease significantly but the light curves at higher frequencies 
remain relatively constant (resulting in a steepening of the spectral index). 
Both the temporal and spectral evolution of V1723 Aql are dramatically 
different from that of the Hubble flow model (shown as dashed lines in Figure 
\ref{v1723aql}). The difference arises from the maximum at early times, but 
our ongoing monitoring indicates that the late time light curve is probably 
settling into a Hubble-flow-like behaviour. There is no indication of a dense 
environment for this source, and neither the low frequency flux density limits 
nor the spectral indices are consistent with a significant non-thermal 
component. A model similar to that developed by \citet{tay87} might apply in 
the case of V1723 Aql, if the absorbing screen is less dense and does not play 
a significant role in steepening the spectral index. The Swift XRT detection 
of hard X-ray emission from V1723~Aql around the Day $\sim$50 bump indicates 
the presence of shock. Further follow up observations, as well as more 
detailed analysis and modeling for this source is in progress (Weston et al. 
in preparation).

Although we are currently limited by a sample size of two, the comparison of 
V1723 Aql and QU Vul demonstrates diversity of early time radio bumps, both in 
rise/decline times and radio spectral indices. This diversity underscores the 
potential richness of radio light curves at early times, and it remains to be 
seen if the same underlying physical process can explain the full range of 
observations.

\subsection{Delayed rises of radio light curves}

Because the radio light curve traces the mass and extent of the ionized nova 
ejecta, they are an elegant tracer of \emph{when} the mass is actually 
expelled. While most historical light curves of novae in the radio are 
consistent with the ejection of mass around the time of optical discovery, our 
recent VLA data on T~Pyx is an intriguing counter-example.

Our multi-frequency light curves of the 2011 outburst of T~Pyx started rising 
surprisingly late, about 80 days after the optical outburst (\citealt{tpyx0}, 
\citealt{tpyx1}, Nelson et al. 2012d). The rise is wholly inconsistent 
with expulsion at the time of optical discovery, and instead implies that the 
gaseous envelope was in rough hydrostatic equilibrium for a couple months, 
until it suddenly started to expand. The time of radio rise also coincides 
with the time of a sharp decline in the optical light curve, which had been 
experiencing plateau-like behaviour around maximum prior to Day $\sim$80. This 
is as expected if the optical photosphere begins to recede when the ejecta 
begin bulk expansion. Such a delayed discrete episode of mass loss is 
unprecedented in radio observations of novae, although previous observations 
may have had limited sensitivity to small delays in ejection because of the 
poor early-time coverage typical in radio light curves. Optical light curves 
of novae do regularly show plateaus of various durations, but this behaviour 
was poorly understood. For the first time, in T~Pyx, our radio observations 
may clearly link plateaus with delayed ejection. In the future, we plan to 
test if pre-maximum halts \citep[as seen in the optical,][]{hou10} and other 
plateaus always coincide with delayed mass ejection, as traced by radio light 
curves.

\subsection{Testing models of mass ejection in novae}

Features such as multi-phase and delayed ejections clearly show that richly 
varied physical processes are instrumental in ejecting material from the 
surface of the WD. The next step is to test, in detail, if this clearly 
observed complexity in mass ejection is consistent with theoretical 
predictions. 

While complex multi-phase ejection histories are predicted by models that 
combine hydrodynamics and networks of nuclear reaction rates to simulate nova 
explosions \citep{pri86}, previous work has focused on the simplest modeled 
quantities (e.g., total ejecta mass, maximum velocity). The extensive grid of 
models used to predict these simple quantities \citep[e.g.,][]{yar05} also 
produces detailed mass-loss histories for simulated novae, but these results 
remain unpublished (M.~Shara 2012, private communication). The E-Nova Project 
plans to carry out a detailed comparison between model predictions of mass 
loss histories from novae and modern multi-wavelength data, with radio 
observations leading the charge and tracing the bulk of the ejected mass.

Radio observations can produce some of the highest resolution images in all of 
astronomy, and high-resolution arrays like eMERLIN, VLBA, and VLA in A 
configuration are all undergoing dramatic upgrades. In the future, our 
observational capabilities promise to keep pace with advances in theory 
brought on by multi-dimensional simulations of novae, which are only in their 
infancy but already revealing new insights into long-standing problems in our 
understanding of nova explosions \citep{cas11}.

\section{The effect of environment on radio light curves}\label{nova:em13}

The environments surrounding novae can also impact the evolution of radio 
emission from these explosions and affect our ability to derive fundamental 
explosion parameters from nova light curves. At the same time, radio emission 
from ``embedded'' novae (surrounded by dense environments) can provide rich 
insights into the nature of circumstellar material in binary systems. If, as 
proposed by \citet{wil10}, significant reservoirs of circumbinary material 
are commonly present around novae, interaction between the ejecta and this 
material should modify the radio light curve considerably.

When nova ejecta interact with a dense environment, a strong shock should be 
formed, which in turn may give rise to a synchrotron emission component. In a 
few cases, like RS Oph \citep{hje86,tay89,obr06,rup08,sok08} and GK Per 
\citep{sea89,anu05}, synchrotron emission is found to be the significant, or 
sometimes even the dominant, emission component. The Fermi detection of the 
recent nova in V407 Cyg revealed that shock interaction with a dense 
environment can also lead to GeV gamma ray emission in novae \citep{abd10}. 
Recently, the Fermi collaboration reported another possible gamma ray 
transient associated with a nova \citep{che12}. We have obtained early time 
radio data for this source, Nova Sco 2012, and our preliminary analysis 
reveals a spectral slope that is compatible with non-thermal emission, as 
expected if a strong shock interaction is taking place \citep{asco2012}.

Interactions between nova ejecta and circumbinary material may also give rise 
to thermal radiation produced by the ionization of circumbinary material by 
the nova outburst. From our VLA observations of the 2010 nova in the symbiotic 
binary V407~Cyg, we see clear and strong effects of the dense wind from the 
Mira giant companion in the radio light curve. More than two years of VLA 
monitoring shows that the radio evolution of V407~Cyg can not be reconciled 
with the standard model of expanding thermal ejecta. The bright radio flux 
densities would require a massive ejection ($10^{-5} - 10^{-4} M_\odot$), 
directly in conflict with other observations suggesting a low ejecta mass 
($10^{-7} - 10^{-6} M_\odot$; \citealt{mun11,sch11,nel12}). Even with massive 
ejecta, the observed spectral indices are inconsistent with thermal nova 
ejecta. For this source, we developed an alternative detailed model of thermal 
radio emission from the surrounding dense Mira wind which was ionized by the 
nova outburst. Our data suggest that this mechanism can produce the observed 
radio luminosity consistently. The rising part of the radio light curve is due 
to the increasing ionization of the wind, whereas the nova shock from within 
causes the declining part. There is no evidence of additional emission 
components, from either thermal nova ejecta or synchrotron shocks; these 
components were likely present but hidden behind the absorbing screen of the 
ionized Mira wind. Our VLA observations and detailed modeling of the radio 
emission are presented in \citet{cho12}.

While RS~Oph, GK~Per, and V407~Cyg are clearly extreme examples of dense 
environments around novae, it is likely that all novae lie on a continuum of 
environmental densities. Other more ``classical'' novae may show subtle yet 
detectable effects of circumbinary material in radio light curves; the more 
extreme embedded novae are useful for demonstrating and calibrating the 
observational effects of such interaction.

\section{Concluding remarks}\label{nova:concl}

With radio and millimeter astronomy on the cusp of enjoying many new exciting 
facilities, radio studies of novae are undergoing a renaissance. With the 
upgraded VLA, we have started detecting many unforeseen complexities in radio 
light curves, which demand refinement of models of radio emission from novae 
and provide opportunities for testing models of the nova explosions 
themselves. From the historical framework developed by R.~Hjellming, 
E.~Seaquist, A.~R.~Taylor, and others, it is clear that radio observations 
provide unique insights into nova explosions, because they trace thermal 
free-free emission, and by extension, the bulk of the ejected mass. Radio 
thermal optical depth evolves on very different timescales, as compared with 
optical observations. In addition, radio observations are not subject to the 
many complex opacity and line effects that optical observations both benefit 
and suffer from. This means that not only can radio observations be much more 
convenient to obtain and interpret, but they can also probe how the ejecta 
profile and dynamic mass loss evolve with time. Continuing our systematic 
radio monitoring with complementary multi-wavelength campaigns, and improving 
the models as demanded by the data, we are confident that radio observations 
of novae will illuminate the many multi-wavelength complexities observed in 
novae and test models of nova  explosions.

\section*{Acknowledgements}

N.~Roy and L.~Chomiuk are Jansky Fellows of the National Radio Astronomy 
Observatory. T.~Johnson is a summer student at the National Radio Astronomy 
Observatory. The National Radio Astronomy Observatory is a facility of the 
National Science Foundation operated under cooperative agreement by Associated 
Universities, Inc.


\label{lastpage}
\end{document}